\documentclass{mem}
\usepackage{natbib}
\usepackage{txfonts}
\usepackage{balance}
\usepackage{graphicx}
\usepackage[a4paper]{hyperref}
\idline{75}{282}
\begin{document}
\def\teff{$T\rm_{eff }$}
\def\kms{$\mathrm {km s}^{-1}$}

\title{Radio relics in the MareNostrum Universe}

   \subtitle{}

\author{
S. E. \,Nuza\inst{1} 
\and M. \,Hoeft\inst{2}
\and S. \,Gottl\"ober\inst{1}
\and R. J. \,van Weeren\inst{3}
\and G. \,Yepes\inst{4}
          }	

  \offprints{S. E. Nuza}

\institute{
Astrophysikalisches Institut Potsdam, An der Sternwarte 16, 14482 Potsdam, Germany\\
\email{snuza@aip.de}
\and
Th\"uringer Landessternwarte, Sternwarte 5, 07778 Tautenburg, Germany
\and
Leiden Observatory, Leiden University, P.O. Box 9513, NL-2300 RA Leiden, Netherlands
\and
Grupo de Astrof\'{\i}sica, Universidad Aut\'onoma de Madrid, Cantoblanco, 28039 Madrid, Spain
}

\authorrunning{Nuza et al. }

\titlerunning{Radio relics number counts}

\abstract{
We identify shocked gas in simulated galaxy clusters extracted from the MareNostrum Universe simulation 
\citep[][]{2006astro.ph..8289G} assuming that shock waves are regions of electron acceleration. 
We perform flux number counts within the framework of the non-thermal emission model developed by \citet{2008MNRAS.391.1511H}. 
Results are presented at two different observing frequencies, i.e. 1.4 GHz and 120 MHz, posing interesting 
constraints for LOFAR and upcoming radio telescopes. 
}
\maketitle{}

\section{Introduction}

Radio relics are elongated structures located in the outskirts 
of galaxy clusters. Seemingly, these radio features are produced during 
cluster merger events as a result of the formation of shock waves, which 
are believed to be ideal sites for electron acceleration. 
In the presence of magnetic fields accelerated electron populations 
are capable of producing the observed synchroton emission. 

During the last years, the number of observed relics has dramatically increased 
due to the spectacular improvement in the instruments sensitivity. In particular, the 
system found in the galaxy cluster CIZA~J2242.8+5301 is one of the most spectacular relics 
known to date, giving strong support to the previously mentioned astrophysical 
scenario \citep{2010Sci...330..347V}.

In this work, we estimate the amount of diffuse radio emission produced in 
simulated relics using a synthetic galaxy cluster sample extracted from a cosmological 
simulation box. After identifying shocks in the simulation we apply a radio emission 
model based on the diffusive shock acceleration (DSA) process and count how many relics 
are observable as a function of redshift to assess expectations for the upcoming LOFAR radio surveys. 

In Sect. \ref{method} we give a brief overview of the method used and 
in Sect. \ref{discussion} we close the contribution with the discussion 
and comment about future work. 

\section{Method}
\label{method}

We select the 500 most massive clusters in the MareNostrum Universe 
simulation \citep[][]{2006astro.ph..8289G}, a non-radiative smoothed particle hydrodynamics 
cosmological run with $2\times1024^3$ gas and dark matter particles, at 3 different 
cosmic times ($z=0,0.5$ and $1$). 

Shock waves in the simulated clusters are identified in order to compute 
the non-thermal emission as a function of Mach number, magnetic field and thermodynamical state of the 
post-shock region. We assume that the magnetic field in the shocks scales with gas density 
simply obeying flux conservation, i.e. $B=B_0\times(n_{\rm e}/n_0)^{2/3}$, where 
$n_{\rm e}$ is the electron density and $n_0$ and $B_0$ are two reference values chosen to be 
$10^{-4}$ cm$^{-3}$ and 0.1 $\mu$G respectively. In our model electrons are accelerated by means of the DSA mechanism 
whereas they cool down due to synchroton looses and inverse compton scattering with the cosmic microwave background. 
For more details about the shock finding technique and the radio model used see \citet{2008MNRAS.391.1511H}. 

For each simulated cluster we estimate the radio emission produced in shocks inside a sphere of size $\sim2\times R_{\rm vir}$ 
with $R_{\rm vir}$ being the virial radius. In order to estimate the total non-thermal emission for less massive 
haloes at each redshift we construct the radio power distribution for relics in clusters with different masses and extrapolate 
the obtained trend. 
From this information we evaluate the probability of finding radio relics for a given cluster mass, 
radio power and cosmic time. To compute the radio relic luminosity function (RRLF) as a function of redshift 
we convolve this probability with the halo mass function given by \citet{2002MNRAS.329...61S}. Finally, using the 
RRLF it is possible to perform the flux number counts. 

\begin{figure}
\begin{center}
{\includegraphics[width=65mm]{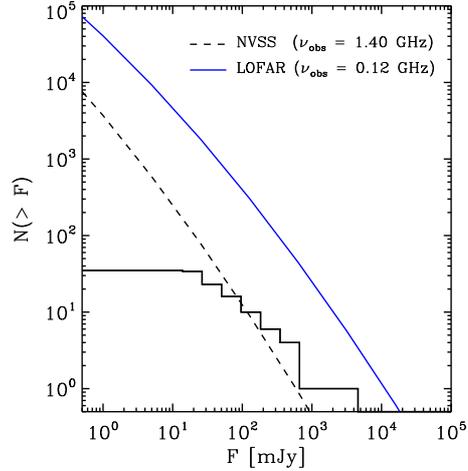}}
\caption{\footnotesize Cumulative number of observed relics as a 
function of radio flux. The histogram shows the observed NVSS relic 
sample ($\delta > -40^{\circ}$) as a reference. 
Also shown are models at 1.4 GHz (NVSS) 
and 120 MHz (LOFAR {\it Tier 1}).
}
\label{fig}
\end{center}
\end{figure}

\section{Discussion}
\label{discussion}

Figure \ref{fig} show the model predictions for the observed relic cumulative number 
per logarithmic radio flux at 1.4 GHz (normalized to the NVSS sample at 100 mJy) 
and 120 MHz (corresponding to the LOFAR {\it Tier 1} survey). It can be seen that 
LOFAR will be able to observe thousands of relics. The actual number will 
depend on the final sensitivity of the survey and on the details of the shock magnetic 
field involved. We will present a comprehensive analysis using different model assumptions 
in a forthcoming paper (Nuza et al. 2011, in preparation).

\begin{acknowledgements}
S.E.N. thanks DFG and Chiara Ferrari for financial support. 
\end{acknowledgements}

\bibliographystyle{aa}
\bibliography{Nuza_proc_nice_2010}

\end{document}